\documentclass[11pt,a4paper]{article}
\usepackage[hyperref]{acl2021}
\usepackage{times}
\usepackage{latexsym}
\usepackage{xcolor} 
\usepackage{graphicx}
\usepackage{comment}
\usepackage{amssymb}
\usepackage{multirow}

\usepackage{array}
\newcolumntype{H}{>{\setbox0=\hbox\bgroup}c<{\egroup}@{}}
\usepackage{microtype}

\aclfinalcopy 

\title{Speech Technology for Everyone: Automatic Speech Recognition for Non-Native English with Transfer Learning}

\author{Toshiko Shibano$^*$ ~~ Xinyi Zhang$^*$ ~~ Mia Taige Li$^*$ ~~ Haejin Cho$^*$ \\ \textbf{Peter Sullivan$^*$ ~~ Muhammad Abdul-Mageed\thanks{~~All authors contributed equally.} }\\
The University of British Columbia \\
\texttt{ \small{tshibano@student.ubc.ca, \{jeremyzxy0803, mia.taige.li, haejin2909\}@gmail.com,}}\\ \texttt{ \small{\ prsull@student.ubc.ca, muhammad.mageed@ubc.ca}}\\

}

\date{}

\begin{document}
\maketitle
\begin{abstract}
To address the performance gap of English ASR models on L2 English speakers, we evaluate fine-tuning of pretrained wav2vec 2.0 models \cite{baevski2020wav2vec,xu2021self} on L2-ARCTIC, a non-native English speech corpus \cite{zhao2018l2} under different training settings. We compare \textbf{(a)} models trained with a combination of diverse accents to ones trained with only specific accents and \textbf{(b)} results from different single-accent models. Our experiments demonstrate the promise of developing ASR models for non-native English speakers, even with small amounts of L2 training data and even without a language model. Our models also excel in the zero-shot setting where we train on multiple L2 datasets and test on a blind L2 test set. 


\end{abstract}

\section{Introduction}\label{sec:intro} 

Although non-native (L2) English speakers outnumber native (L1) English speakers \cite{crystal2003english}, major challenges contribute to a gap between performance of ASR systems on L2 speech, mainly due to the influence of L1 pronunciation on the learned language, and the lack of annotated L2 speech data~\cite{radzikowski2021accent, viglino2019end}. To meet these challenges, previous studies have exhibited two distinct approaches. The first is to make L2 speech representations more closely match those of L1 speech~\cite{radzikowski2021accent}. The second approach leverages L2 speech data to improve model robustness. Due to L2 data scarcity, and hence the challenge of training L2 models from scratch, this second approach necessitates employment of transfer learning or domain adaptation~\cite{shi2021accented,sun2018domain}.

\begin{figure}[h]
\centering{\includegraphics[width=\linewidth]{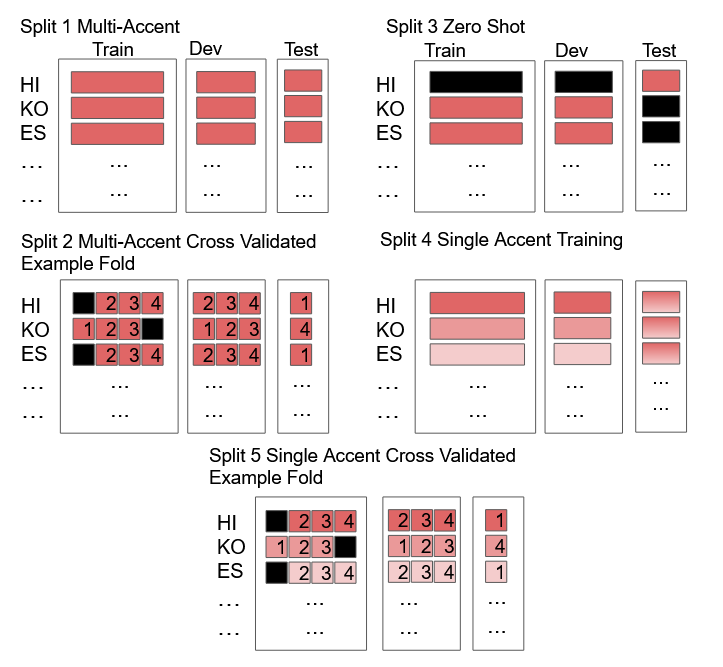}}
\caption{\label{fig:splits}
The various data splits we use in our experiments. Shade represents a different run of our training, with the gradient blocks in Split 4 being present in all runs. For cross validation splits, we show a single fold as an example, where number indicates the participants included. 
}
\end{figure}


State-of-the-art ASR models based on unsupervised/self-supervised pre-training such as wav2vec ~\cite{schneider2019wav2vec} and wav2vec 2.0~\cite{baevski2020wav2vec}\footnote{Although sometimes referred to as `unsupervised', these models employ a self-supervised objective.} offer a tantalizing starting point for applying the second approach we list above, especially due to their strong performance on ASR even without a language model. However, challenges remain in identifying how best to apply models such as wav2vec 2.0 in L2 fine-tuning scenarios. For this reason, our objective in the current work is to investigate a rich set of conditions under which we can fine-tune ASR models for optimal L2 performance. More concretely, we attempt to achieve the following:

\begin{enumerate}
    \item Evaluate fine-tuning strategies for adapting pre-trained L1 English ASR models to L2 English;
    \item Explore impact of non-native (L2)  accents on performance of these fine-tuned ASR models, comparing multi-accent training to single-accent training; and
    \item Quantify the impact of L2 fine-tuning on model performance for L1 English speech recognition.
\end{enumerate}

Although external language models are often used in improving ASR performance~\cite{nakatani2019improving, xu2020independent}, models trained with great quantities of data can potentially internalize this linguistic information ~\cite{graves2014towards}. In particular, some of the wav2vec 2.0 models perform nearly as well with and without a language model on difficult speech such as LibriSpeech Test-Other ~\cite{xu2021self}. We thus use this robust pre-trained model as our starting point, and carry out our work without use of an external language model to see if this performance is retained through the fine-tuning process. 

The rest of the paper is organized as follows: Section~\ref{sec:lit} is an overview of related works. We describe our data in Section~\ref{sec:data}. Section~\ref{sec:exp} is about our experiments and results. We conclude in Section~\ref{sec:con}.

\section{Related Work}\label{sec:lit}


Because of the difficulty in linguistically annotating corpora for Hidden Markov Model (HMM)-based ASR ~\cite{graves2014towards}, researchers have broadly embraced End-to-End (E2E) deep learning architectures either based on Connectionist Temporal Classification (CTC) ~\cite{graves2006connectionist,graves2014towards}, Attention ~\cite{chorowski2015attention,chan2016listen, gulati2020conformer}, or hybrids of the two ~\cite{watanabe2017hybrid,wang2020transformer}. Recent efforts inspired by work such as BERT ~\cite{devlin-etal-2019-bert} have improved on these purely supervised learning baselines through self-supervised pre-training ~\cite{schneider2019wav2vec,baevski2019vq, baevski2020wav2vec} and self-training ~\cite{xu2021self}. These self-supervised wav2vec models represent one line of research in speech representation. Other works include models similar to wav2vec that also use a contrastive loss~\cite{oord2018representation}, models using an autoregressive loss function~\cite{ling2020deep,chung2019unsupervised}, as well as models using a masked language model closer to the original BERT ~\cite{liu2020mockingjay}.

With these efforts, ASR technologies for native languages have evolved significantly. However, we still observe problems in many applications. In particular, several researchers have emphasized how performance of ASR models drops when the input speech is from non-native speakers whose native languages are different from the models' target languages~\cite{radzikowski2021accent,livescu2000lexical,wang2003comparison,ping2008automatic}. For systems developed for English ASR, this can be a real issue. The reason, as observed earlier, is that large populations of English language speakers are non-native~\cite{crystal2003english}. In line with this argument,~\newcite{ping2008automatic}, for example, pointed out the necessity to improve speech recognition technology for L2 speakers given that many people speak more than one language for economic and social reasons, especially considering human migration is becoming more common these days. It is hoped that continued efforts aiming at improving ASR for non-native speakers will eventually lead to improved results for many as voice recognition technology becomes increasingly pervasive in our daily lives~\cite{ping2008automatic}.

As we explained in Section~\ref{sec:intro}, there are two distinct approaches to improve current ASR performance on L2 speech: 1) accent conversion as an extension to the active area of research of voice conversion; and 2) incorporation of L2 speech data, which is often limited in quantity and quality, during the model training process. 

The first approach takes inspiration from voice conversion, but instead of focusing on modifying the pitch, it modifies the pronunciation to reduce accents. Additionally, voice conversion models aim to generate results that are speaker-dependent, while accent conversion models deal with generalizing accents from a group of speakers, hence being speaker-independent. With this approach, the resulting model can be used as a pre-processing step to remove accents in the data prior to feeding these data into an ASR model.~\newcite{bearman2017accent} adopt this approach but focus on L1 English accents, while ~\newcite{radzikowski2021accent} work on L2 English accents with speakers' L1 being Japanese. ~\newcite{liu2020end} took a step further and turned Hindi-accented English to native American English without utilizing native utterances.

The second approach often employs techniques such as domain adversarial training and transfer learning in order to utilize as much available accented speech data as possible. Domain adversarial training (DAT) is a popular approach as it encourages models to learn accent-invariant features~\cite{sun2018domain, hou2019domain, hu2021redat}. Transfer learning is another popular approach in L2 speech recognition, as it possibly allows a model to gain knowledge from both the base task and the new task, even when the new task has limited data~\cite{matassoni2018non, das2021best, shi2021accented}. In the Accented English Speech Recognition Challenge 2020 (AESRC2020), many teams utilize transfer learning to tackle the L2 accent recognition task~\cite{shi2021accented}. In a recent work, ~\newcite{das2021best} combine both DAT and transfer learning to achieve robust accented speech recognition performance. We now introduce our data.

\section{Data}\label{sec:data}

\subsection{Corpus Information}
We choose \textbf{L2-ARCTIC}, a non-native English speech corpus~\cite{zhao2018l2}, for L2 fine-tuning. The recordings are from 24 non-native speakers of English with a total of six different L1s, and each of the L1s consists of two female speakers and two male speakers. The L1s we use for our experiments are Arabic (AR), Hindi (HI), Korean (KO), Mandarin (ZH), Spanish (ES), and Vietnamese (VI). Because L2-ARCTIC is based on the original L1 English corpus, CMU ARCTIC~\cite{kominek2003cmu} (henceforth \textbf{L1-ARCTIC}, for simplicity), we can easily evaluate performance from fine-tuning on same-domain L1 data.

 Each speaker in L2-ARCTIC contributed approximately one hour of phonetically-balanced read speech based on the L1-ARCTIC prompts, which consist of carefully selected sentences ($1,132$ sentence prompts) from Project Gutenberg \cite{kominek2003cmu}. We note this, as the pretrained wav2vec 2.0 model we use was first pre-trained on LibriSpeech\footnote{http://www.openslr.org/12/}~\cite{panayotov2015librispeech} and then self-trained on Libri-Light\footnote{https://github.com/facebookresearch/libri-light}~\cite{kahn2020libri}.  Both corpora rely on audiobooks from the LibriVox project,\footnote{https://librivox.org} much of which comes from Project Gutenberg.\footnote{http://www.gutenberg.org} This minimizes discrepancies between domains of the text.

We also evaluate our fine-tuned models on \textbf{1) LibriSpeech} to compare the fine-tuning with the original performance of self-trained wav2vec 2.0 Large (LV-60) model~\cite{xu2021self}, which we will refer to as \textit{Wav2Vec 2.0-ST}. In addition, we evaluate on \textbf{2) L1-ARCTIC}, identical to our L2-ARCTIC corpus but spoken by four native US English speakers, allowing us to identify any degradation in performance on L1 speech. Each of L1-ARCTIC speakers' datasets contain approximately the same number of utterances ($n=\sim1,132*4$) as each of L2-ARCTIC speakers' datasets.\par
For the purpose of our experiments, we define \textit{native (L1) accents} as those represented in the LibriSpeech and L1-ARCTIC, and \textit{non-native (L2) accents} as those represented in L2-ARCTIC.

\begin{table*}[]
\centering
\begin{tabular}{llcclcc}
\hline
 &  & \multicolumn{2}{c}{\textbf{Accent dependency}} &  & \multicolumn{2}{c}{\textbf{Speaker dependency}} \\ \cline{3-4} \cline{6-7} 
 &  & \multicolumn{1}{l}{\textbf{Dependent}} & \multicolumn{1}{l}{\textbf{Independent}} &  & \multicolumn{1}{l}{\textbf{Dependent}} & \multicolumn{1}{l}{\textbf{Independent}} \\ \hline
\textbf{Multi-accent} & Model-1 (Split 1) & x &  &  & x &  \\
 & Model-2 (Split 2) & x &  &  &  & x \\
 & Model-3 (Split 3) &  & x &  &  & x \\ \hline
\textbf{Single-accent} & Model-4 (Split 4) & x & x &  & x & x \\
 & Model-5 (Split 5) & x &  &  &  & x \\ \hline
\end{tabular}
\caption{Summary of data splits, fine-tuning, and evaluation setups.}
\label{table:model_summary}
\end{table*}


\subsection{Data Splits}
For both L2-ARCTIC and L1-ARCTIC, we split the data into three distinct Train, Dev, and Test sets with an $80$:$10$:$10$ ratio. Importantly, we ensure there is \textit{no overlap between utterances}. For  L2-ARCTIC, we split the data across the following settings (see Fig. \ref{fig:splits}).

\begin{itemize}
    \item  \textbf{Split-1} \textit{(speaker-dependent, multi-accent split)}: All speakers from all accents in the Train set are also included in the Dev and Test sets; however, no utterances are shared between Train, Dev, and Test.
    
    \item  \textbf{Split-2} \textit{(speaker-independent cross-validation splits with multiple accents)}: A speaker from each accent\footnote{We use the term `accent' here to loosely refer to variation in speakers with L1 other than English.} is removed from the Train and Dev sets, but other speakers with the same accent remain in the Train and Dev sets. 

    \item  \textbf{Split-3} \textit{(speaker-independent zero-shot splits with multiple accents)}: All speakers from one of the accents are entirely removed from the Train and Dev sets. The removed speakers are included in Test. 
 
    \item   \textbf{Split-4} \textit{(all-speaker, single-accent split)}: Speakers are broken down by accents (six accents in total) and all speakers in a given accent are split into the Train, Dev, and Test sets (3 data splits x 6 accents).
 
    \item   \textbf{Split-5} \textit{(speaker-independent cross-validation splits with single accent)}: One speaker in each accent is removed from the Train and Dev sets, but the other speakers with the same accent remain in the Train and Dev sets. As there are four speakers per accent, four splits are created for each accent, which are further split into the Train, Dev, and Test sets (3 data splits x 6 accents x 4 speakers).

\end{itemize}

\section{Experiments}\label{sec:exp}

 For all our wav2vec 2.0 models, we use Fairseq~\footnote{\url{https://github.com/pytorch/fairseq}} fine-tuning default settings as a reference and convert the hyper-parameters to align with Huggingface's implementation. We train each model with three random seeds and take average over three WERs, one each from the three seeds.

\subsection{Model Architecture, Fine-tuning, Baselines, and Evaluation}
For our model development, we use the wav2vec 2.0 architecture~\cite{baevski2020wav2vec} which is composed of a multi-layer convolutional neural network feature extractor and a Transformer context network. It takes in raw audio and converts it into representations of the input sequence.  The encoder consists of multiple blocks of temporal convolution followed by a layer normalization and a GELU activation function. The relative positional embedding in the Transformer is accomplished by a convolutional layer.\par 

Fine-tuning of pre-trained wav2vec 2.0 is performed with CTC and the transcriptions of the audio segments. For each model, we identify the optimal hyper-parameters on the respective Dev set. We choose hyper-parameters as follows: For \texttt{mask\_feature\_prob}, we pick from \textit{\{0.25, 0.5\}}, for \texttt{mask\_feature\_length}, we choose from \textit{\{15, 30\}}, for \texttt{mask\_time\_prob} we use \textit{\{0.5, 0.75\}}, and a batch size of $16$. To mimic the tri-state learning rate schedule~\cite{baevski2020wav2vec}, we set different learning rates for different stages:  warm-up (1e-5, 3e-5), constant stage (1e-5, 3e-5), and decay (1e-5, 3e-5, 5e-6). The decay stage is followed by another constant stage (1e-5, 2e-6, 5e-6) to simulate the Fairseq's fine-tuning configuration. We evaluate all our models in terms of word error rate (WER). \textit{All} our results are the average of three runs, and we use the following baselines:



\begin{itemize}

\item  \textbf{Baseline-I:} Wav2Vec 2.0-ST \cite{xu2021self},\footnote{\url{https://github.com/pytorch/fairseq/tree/master/examples/wav2vec##wav2vec 2.0}} a self-trained version of wav2vec 2.0 \cite{baevski2020wav2vec} exploiting a Transformer large architecture and pre-training on 960 hours of speech data from LibriSpeech~\cite{panayotov2015librispeech}. The self-training is performed on 60K hours of Libri-Light~\cite{kahn2020libri}. We believe this as an already strong baseline. We use the model released via HuggingFace.~\footnote{\url{https://huggingface.co/facebook/wav2vec2-large-960h-lv60-self}}

\item  \textbf{Baseline-II:} This is Wav2Vec 2.0-ST, the same as Baseline-I, fine-tuned on L1-ARCTIC described earlier. The purpose of Baseline-II is to allow for measuring the trade-off of L1 English ASR performance by fine-tuning the English pre-trained model on L2 accents. 
\end{itemize}


\begin{table*}[ht]
\centering
\begin{tabular}{llrrlrrlrrlrr}
\hline
 &  & \multicolumn{2}{c}{\textbf{L2-ARCTIC}} &  & \multicolumn{2}{c}{\textbf{L1-ARCTIC}} &  & \multicolumn{2}{c}{\textbf{LS\textsubscript{dev}}} &  & \multicolumn{2}{c}{\textbf{LS\textsubscript{test}}} \\ \cline{3-4} \cline{6-7} \cline{9-10} \cline{12-13} 
\textbf{Model} &  & \multicolumn{1}{r}{\textbf{Dev}} & \multicolumn{1}{r}{\textbf{Test}} & \multicolumn{1}{r}{} & \multicolumn{1}{r}{\textbf{Dev}} & \multicolumn{1}{r}{\textbf{Test}} & \multicolumn{1}{r}{} & \multicolumn{1}{r}{\textbf{Clean}} & \multicolumn{1}{r}{\textbf{Other}} & \multicolumn{1}{r}{} & \multicolumn{1}{r}{\textbf{Clean}} & \multicolumn{1}{r}{\textbf{Other}} \\ \hline
Baseline-I &  & 13.47 & 12.47 &  & 2.30 & 2.23 &  & \textbf{1.69} & \textbf{3.55} &  & \textbf{1.86} & \textbf{3.89} \\
Baseline-II &  & 17.29 & 15.95 &  & \textbf{1.26} & \textbf{1.30} &  & 2.19 & 5.13 &  & 2.32 & 5.00 \\
Model-1 &  & \textbf{9.78} & \textbf{9.27} &  & 1.94 & 1.86 &  & 2.75 & 5.55 &  & 2.82 & 6.36 \\ \hline
\end{tabular}
\caption{\label{model-1} 
    Model-1 performance in word error rate (WER) (lower is better) on non-native accents (L2-ARCTIC) and native accents (L1-ARCTIC, LS\textsubscript{dev} and LS\textsubscript{test}). Baseline-I and Baseline-II are reported on the same Dev and Test sets of each corpus for comparison.
    }
\end{table*}

\begin{figure}[h]
\centering{\includegraphics[scale=.7]{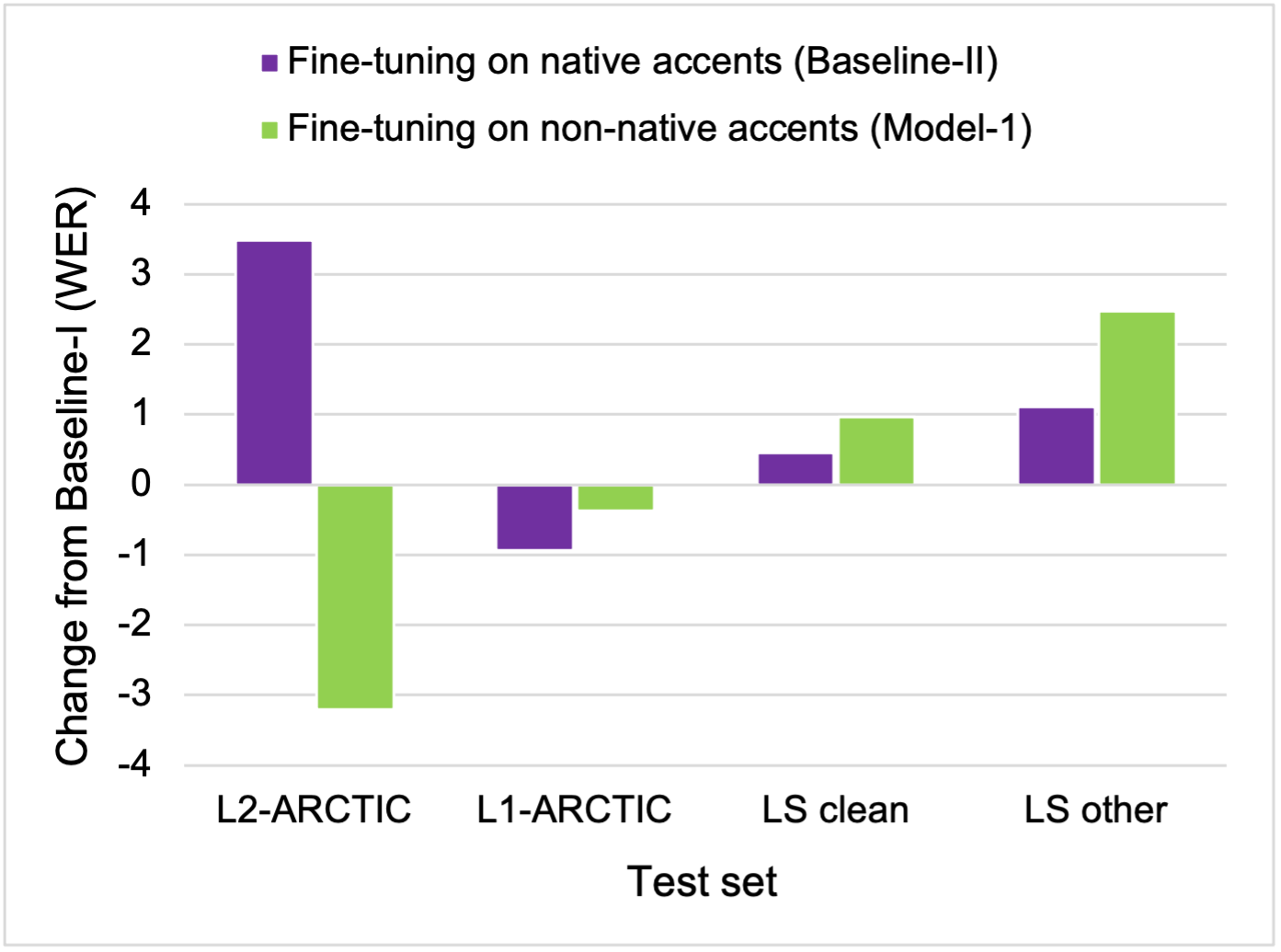}}
\caption{\label{fig:trade-off}
Trade-offs of fine-tuning on native accents (Baseline-II) vs. non-native accents (Model-1). As we evaluate model accuracy by error rate, the bars extending into the negative values mean that the model gains accuracy by fine-tuning. 
}
\end{figure}

\begin{table}[h]
\begin{tabular}{llrrlrr}
\hline
 &  & \multicolumn{2}{c}{\textbf{Dev\textsubscript{L2}}} &  & \multicolumn{2}{c}{\textbf{Test\textsubscript{L2}}} \\ \cline{3-4} \cline{6-7} 
\textbf{Model} &  & \multicolumn{1}{r}{\textbf{Mean}} & \multicolumn{1}{r}{\textbf{SD}} &  & \multicolumn{1}{r}{\textbf{Mean}} & \multicolumn{1}{r}{\textbf{SD}} \\ \hline
Baseline-I &  & 13.47 & 0.23 &  & 12.47 & 0.84 \\
Baseline-II &  & 17.29 & 0.41 &  & 15.96 & 1.58 \\
Model-2 &  & \textbf{9.57} & 0.19 &  & \textbf{9.96} & 0.64 \\ \hline
\end{tabular}
\caption{\label{model-2} 
    Model-2 cross validated performance on L2-ARCTIC Dev and Test sets, alongside Baseline-I and Baseline-II performance on the same cross validation splits. Mean refers to the average WER over the four runs and SD refers to the standard deviation.
}
\end{table}

\begin{table*}[h]
\centering
\begin{tabular}{cccccc|Hcc}
\hline
  &  & \textbf{Baseline-I} &  & \textbf{Baseline-II} &  & \multicolumn{3}{c}{\textbf{Model-3}} \\
  \cline{3-3} \cline{5-5} \cline{7-9}
 \textbf{L1\textsubscript{removed}} &  & \textbf{Test\textsubscript{zeroshot}} &  & \textbf{Test\textsubscript{zeroshot}} &  & \textbf{Dev\textsubscript{L2}} & \textbf{Test\textsubscript{zeroshot}} & \textbf{Test\textsubscript{all}} \\
 \hline
 VI &  & 23.30 &  & 28.81 &  & \textbf{7.96} & 18.81 & 9.43\\
 ZH &  & 14.85 &  & 19.32 &  & 9.02 & 12.13 & 9.08\\
 AR &  & 10.95 &  & 14.82 &  & 9.40 & 10.10 & 9.13\\
 ES &  & 10.48 &  & 13.48 &  & 9.38 & 8.89 & \textbf{8.98}\\
 KO &  & 8.18 &  & 10.22 &  & 10.10 & 6.95 & 9.01\\
 HI &  & \textbf{6.93} &  & \textbf{8.93} &  & 10.29 & \textbf{6.67} & 9.11\\
 \hline
\end{tabular}
 \caption{\label{model-3}
     Model-3 setting, where a different accent is removed each run. Test\textsubscript{all} refers to Test of \textit{all} 24 speakers, and Test\textsubscript{zeroshot} refers to Test of those four speakers who have L1\textsubscript{removed} accent. Baseline-I acquires $12.47$ on Test\textsubscript{all}, while Baseline-II acquires $15.95$ on the same test set (i.e., Test\textsubscript{all}).
 }
\end{table*}

\begin{table*}[ht]
\centering
\begin{tabular}{llrlrlrlr|rrr}
\hline
 &  & \textbf{Baseline-I} &  & \textbf{Baseline-II} &  & \textbf{Model-1} &  & \multicolumn{4}{c}{\textbf{Model-4}} \\ \cline{3-3} \cline{5-5} \cline{7-7} \cline{9-12} 
\textbf{L1} &  & \textbf{Test\textsubscript{L2}} &  & \textbf{Test\textsubscript{L2}} &  & \textbf{Test\textsubscript{L2}} &  & \textbf{Test\textsubscript{L2}} & \textbf{Test\textsubscript{L1}} & \textbf{LS\textsubscript{Clean}} & \textbf{LS\textsubscript{Other}} \\ \hline
VI &  & 23.30 &  & 28.81 &  & 15.14 &  & \textbf{12.12} & 2.02 & 3.08 & 6.96 \\
ZH &  & 14.85 &  & 19.32 &  & 11.49 &  & \textbf{8.95} & 1.82 & 2.84 & 6.22 \\
AR &  & 10.95 &  & 14.82 &  & 8.90 &  & \textbf{6.92} & 1.55 & 2.66 & 6.24 \\
ES &  & 10.48 &  & 13.48 &  & 8.92 &  & \textbf{6.68} & 1.56 & 2.53 & 6.11 \\
KO &  & 8.18 &  & 10.22 &  & 6.60 &  & \textbf{4.99} & 1.71 & 2.51 & 5.63 \\
HI &  & 6.93 &  & 8.93 &  & 5.51 &  & \textbf{4.99} & 1.52 & 2.36 & 6.05 \\ \hline
\textbf{Mean} &  & 12.45 &  & 15.93 &  & 9.43 &  & 7.44 & 1.70 & 2.66 & 6.20 \\ \hline
\textbf{SD} &  & 5.97 &  & 7.30 &  & 3.49 &  & 2.72 & 0.20 & 0.26 & 0.43 \\ \hline
\end{tabular}
\caption{Model-4 performance on L2 accent (Test\textsubscript{L2}) and native accent (Test\textsubscript{L1}, LS\textsubscript{Clean}, LS\textsubscript{Other}), compared with Baseline-I, Baseline-II, and Model-1. SD refers to the standard deviation.}
\label{model-4}
\end{table*}


\begin{table*}[h]
\centering
\begin{tabular}{lrrrrrr}
\hline

 & \multicolumn{1}{c}{\textbf{VI}} & \multicolumn{1}{c}{\textbf{ZH}} & \multicolumn{1}{c}{\textbf{AR}} & \multicolumn{1}{c}{\textbf{ES}} & \multicolumn{1}{c}{\textbf{KO}} & \multicolumn{1}{c}{\textbf{HI}} \\ \hline
\textbf{Baseline-I} & 23.30 & 14.85 & 10.95 & 10.48 & 8.18 & 6.93 \\ \hline
\textbf{VI-specific} & 12.12 & 13.62 & 13.01 & 9.95 & 8.55 & 9.62 \\ 
 \texttt{$\Delta$WER } & -11.18 & -1.23 & 2.06 & -0.53 & 0.37 & 2.69 \\
\texttt{$\Delta \%$ } & -48.00 & \textbf{-8.31} & 18.84 & -5.03 & 4.52 & 38.77 \\ \hline
\textbf{ZH-specific} & 20.37 & 8.95 & 11.42 & 9.79 & 6.82 & 10.91 \\ 
\texttt{$\Delta$WER} & -2.93 & -5.90 & 0.47 & -0.69 & -1.36 & 3.98 \\
\texttt{$\Delta \%$} & -12.58 & -39.75 & 4.26 & -6.62 & \textbf{-16.67} & 57.43 \\ \hline
\textbf{AR-specific} & 23.88 & 14.86 & 6.92 & 9.86 & 9.16 & 7.74 \\ 
 \texttt{$\Delta$WER} & 0.58 & 0.01 & -4.03 & -0.62 & 0.98 & 0.81 \\
\texttt{$\Delta \%$} & 2.47 & 0.07 & -36.83 & \textbf{-5.92} & 11.94 & 11.69 \\ \hline
\textbf{ES-specific} & 20.71 & 13.99 & 11.00 & 6.68 & 7.92 & 8.66 \\ 
\texttt{$\Delta$WER} & -2.59 & -0.86 & 0.05 & -3.80 & -0.26 & 1.73 \\
\texttt{$\Delta \%$} & \textbf{-11.13} & -5.81 & 0.43 & -36.23 & -3.22 & 25.01 \\ \hline
\textbf{KO-specific} & 20.07 & 12.12 & 11.66 & 10.04 & 4.99 & 9.09 \\ 
 \texttt{$\Delta$WER} & -3.23 & -2.73 & 0.71 & -0.44 & -3.19 & 2.16 \\
\texttt{$\Delta \%$} & -13.88 & \textbf{-18.38} & 6.45 & -4.23 & -39.04 & 31.17 \\ \hline
\textbf{HI-specific} & 26.18 & 18.39 & 13.51 & 11.90 & 10.72 & 4.99 \\ 
\texttt{$\Delta$WER} & 2.88 & 3.54 & 2.56 & 1.42 & 2.54 & -1.94 \\
\texttt{$\Delta \%$} & \textbf{12.37} & 23.82 & 23.35 & 13.55 & 31.01 & -27.99 \\ \hline
\end{tabular}
\caption{Model-4 performance in the zero-shot setting. Bold fonts represent the accent whose WER drops the most in the zero-shot setting. For example, compared with Baseline-I, the VI-specific fine-tuning not only improves performance on VI (i.e., a drop in WER), but also improves on ZH despite ZH being the unseen accent. One notable pattern is that HI-specific fine-tuning only benefits HI-accented speech recognition while all the other fine-tuning hinder performance on the HI accent.}
\label{model-4 zeroshot}
\end{table*}

\begin{figure}[h]
\centering{\includegraphics[scale=.7]{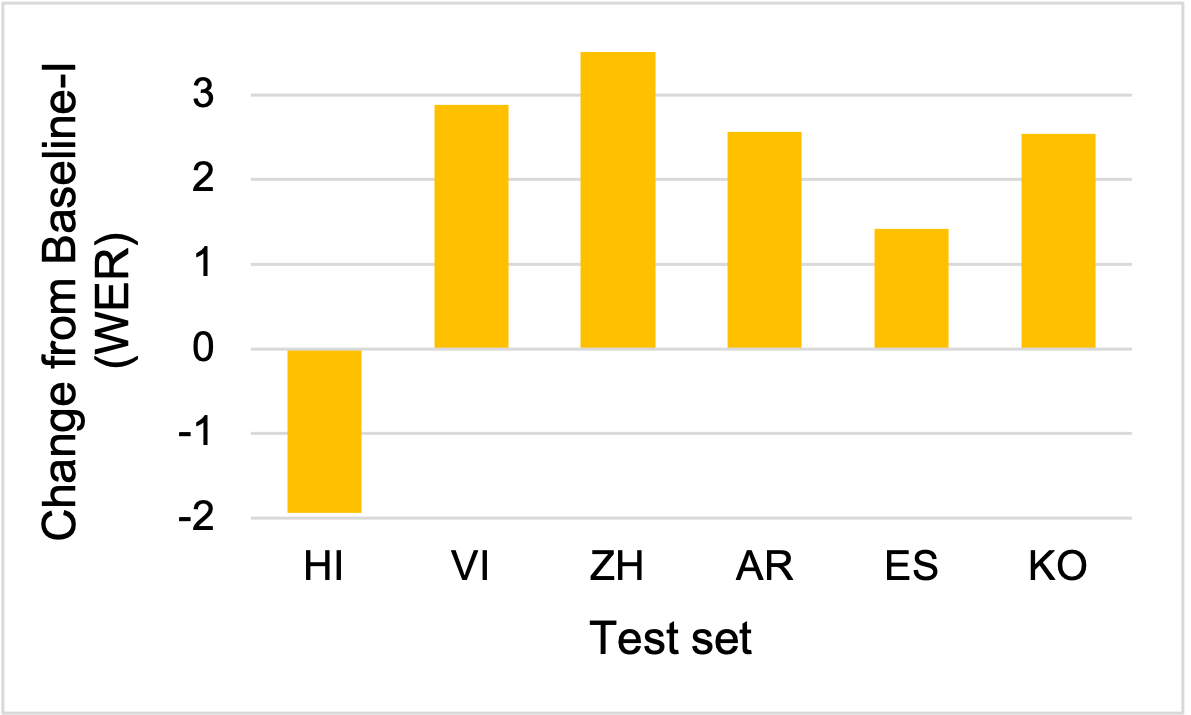}}
\caption{\label{fig:model4-zeroshot1}
HI-specific Model-4 evaluated on individual accents. As we evaluate model accuracy by error rate, the bars extending downwards represent the performance gain by fine-tuning. HI-specific fine-tuning benefits HI but hinders performance on all the other accents.
}
\end{figure}

\begin{figure}[h]
\centering{\includegraphics[scale=.7]{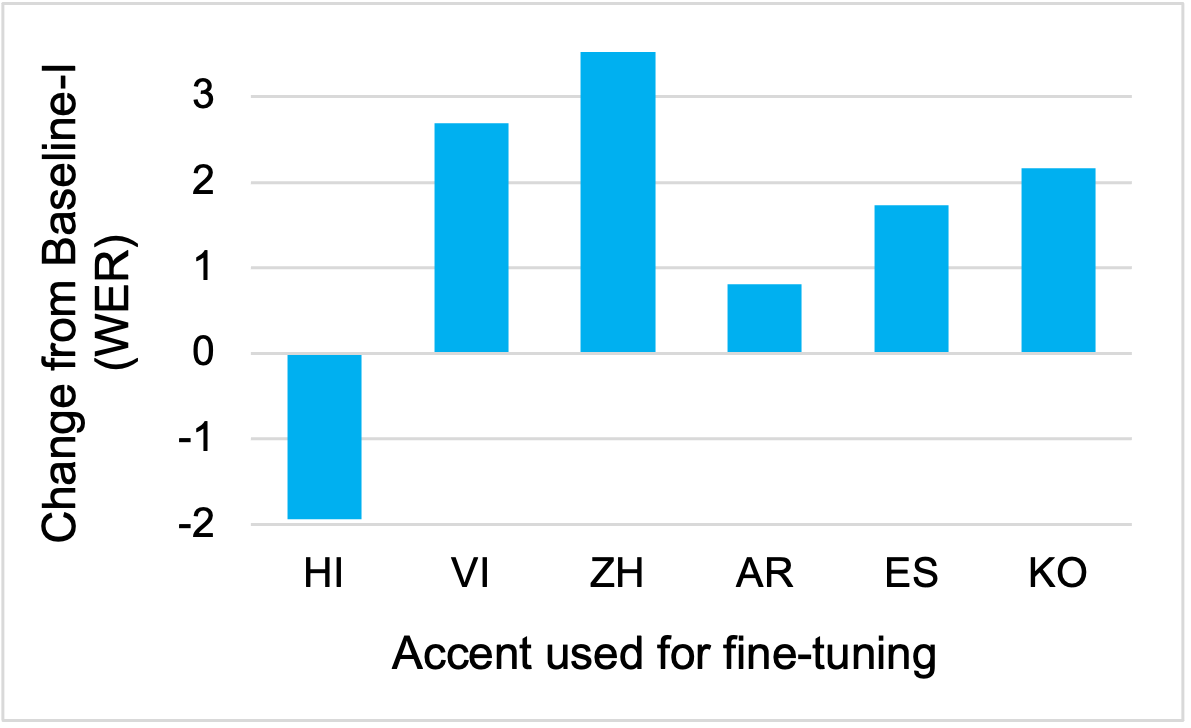}}
\caption{\label{fig:model4-zeroshot2}
Individual Model-4s evaluated on the HI accent. All the bars except HI extend upwards, meaning that all the other single-accent models hinder performance on the HI accent.
}
\end{figure}

\subsection{Multi-Accent Models}

With our multi-accent models, we examine performance using multiple accents during training. We introduce each of our models here, and present the results acquired with each. We provide a summary of our different data splits and models across accent and speaker dependency categories in Table~\ref{table:model_summary}.

     \textbf{Model-1 (speaker- and accent-dependent):}
    The model is fine-tuned with Split-1 data to identify any speaker-dependent training impact, as well as an upper limit on performance. In addition to evaluating on L2-ARCTIC Test, we evaluate on L1-ARCTIC Test and LibriSpeech in order to observe any changes in model performance on L1 English. 
    
    As Table~\ref{model-1} shows, our Model-1 achieves best performance on both Dev and Test of \textbf{L2-ARCTIC} as compared to our two baselines. On Test, our Model-1 acquires $25.66\%$ improvement over our Baseline-I wav2vec 2.0 system on L2-ARCTIC ($9.27$ WER for our model vs. $12.47$ WER for Baseline-I). This gain is not surprising and simply means that a model with access to L2 data for fine-tuning will improve over models fine-tuned with L1 data (Baseline-II, which is fine-tuned on L1-ARCTIC) or not-fine-tuned at all (Baseline-I). Nor is performance on \textbf{L1-ARCTIC} surprising: a model fine-tuned with native data (Baseline-II) outperforms one fine-tuned with accented data (our Model-1), both of which outperform a model without fine-tuning (Baseline-I). These results, however, show that in absence of L1 data, L2 data can be valuable for improving ASR model performance even on L1. For \textbf{LibriSpeech}, Baseline-I, which is trained on LibriSpeech data, outperforms the two fine-tuned models (our Model-1 and Baseline-II). The reason is that these two latter models are fine-tuned on a domain that is different from LibriSpeech. That is, fine-tuning models on out-of-domain data will, and as we see here does, result in deterioration of performance on in-domain data. We also note that our Model-1's performance on LibriSpeech is worse than that of Baseline-II on both the `Clean' (LS\textsubscript{Clean}, native speech under quite recording environments), and `Other' (LS\textsubscript{Other}, both noisy environment and accented recordings),  Dev and Test splits. This may be because LibriSpeech is mostly comprised of L1 data and the greater variability on our L2-ARCTIC Train set (24 non-native speakers in our Model-1 vs. 4 native speakers in Baseline-II).
    


    \textbf{Model-2 (speaker-independent, accent-dependent):} While Model-1 mimics a situation where we have some training data from speakers that we serve (i.e., test on), this is rarely a realistic scenario. We instead switch to a speaker-independent (but still \textit{accent-dependent}) setting, Split-2. We carry out four-fold cross-validation with the 24 speakers in the data, every time using 18 speakers (three speakers per accent) in Train\footnote{We use 10\% of the utterances from these 18 speakers for development (Dev).} and six speakers in Test (one per accent). We report the average of the four folds/runs, along with standard deviation. 
    
    As Table~\ref{model-2} shows, Model-2 performance is consistent with Model-1. Our Model-2 outperforms the two baselines on both Dev and Test, reaching $9.96$ WER on Test compared to $12.47$ for Baseline-I and $15.96$ for Baseline-II. These results demonstrate that fine-tuning with multiple accents improves the accented ASR system without access to test speaker data. 
    

     \textbf{Model-3 (speaker- and accent-independent):} To evaluate performance on \textit{unseen} accents, we adopt a zero-shot strategy by removing one accent at a time from both Train and Dev sets and evaluating on the Test set of the removed accent, Split-3. To evaluate model performance on each accent, we conduct six runs in total with one accent removed at a time. 
    
    As Table~\ref{model-3} shows, fine-tuning on accented speech benefits unseen accents and speakers (Model-3 setting). All the multi-accent, zero-shot models outperform Baseline-I and Baseline-II, which means each of the six accents benefit from other accents through this process of transfer learning. Our results also show that, in absence of in-accent data, some unseen accents are easier for the model than others. For example, on Test\textsubscript{zeroshot}, Vietnamese (VI) is the most challenging (with $18.81$ WER) and Hindi (HI) is the least challenging (with only $6.67$ WER).

\begin{table}[h]
\centering
\begin{tabular}{llrrlrr}
\hline
 &  & \multicolumn{2}{c}{\textbf{Test\textsubscript{all}}} &  & \multicolumn{2}{c}{\textbf{Test\textsubscript{zeroshot-speaker}}} \\ \cline{3-4} \cline{6-7} 
\textbf{L1} &  & \multicolumn{1}{r}{\textbf{Mean}} & \multicolumn{1}{r}{\textbf{SD}} &  & \multicolumn{1}{r}{\textbf{Mean}} & \multicolumn{1}{r}{\textbf{SD}} \\ \hline
VI &  & 12.67 & 0.38 &  & 14.28 & 4.87 \\
ZH &  & 9.65 & 0.31 &  & 11.26 & 3.03 \\
AR &  & 7.28 & 0.29 &  & 8.56 & 2.28 \\
ES &  & 6.95 & 0.26 &  & 7.76 & 3.99 \\
KO &  & 5.22 & 0.18 &  & 5.69 & 2.20 \\
HI &  & 5.27 & 0.11 &  & 5.79 & 1.12 \\ 
\hline
\end{tabular}
\caption{Model-5 performance on L2 accent. Test\textsubscript{all} contains utterances by all speakers within each L1 whereas Test\textsubscript{zeroshot-speaker} contains utterances by a single speaker that is absent in the training phase. Mean refers to the average WER over four folds for each L1, and SD refers to the standard deviation.}
\label{model-5} 
\end{table}

\begin{table*}[h]
\centering
\small    
\begin{tabular}{c|l}

\hline
\multicolumn{1}{l|}{\textbf{Model}} & \multicolumn{1}{c}{\textbf{Model output}} \\
\hline
Ref & at lake linderman i had one canoe very good peterborough canoe \\
\hline
\multirow{2}{*}{VI} & at LAY LINDEMAN i had one canoe very good PETERBORROUG CANOES \\
 & A lake LNDER MAN i had one canoe very good BIET OF ROCK canoe \\
\hline
\multirow{2}{*}{ZH} & at lake LINGERMAN i had ONCE canoe very good PETERBROUGH canoe\\
 & at lake LINERMAN i had one canoe very good PETERE BROUGHTA canoe \\
\hline
\multirow{2}{*}{AR} & at lake LUNDERBOGH i had one canoe very good BITTERBOROUGH canoe \\
 & at lake LUNDERMAN i had one canoe very good BETTER BORT canoe \\
\hline
\multirow{2}{*}{ES} & at lake linderman i had one canoe a very good PETERBOURN canoe \\
 & at lake linderman i had ONCE canoe very good PIERREBOROUGH canoe \\
\hline
\multirow{2}{*}{KO} & at lake linderman i had one canoe very good peterborough canoe\\
 & at lake LINDEMAN i had ONCE canoe very good PITTEBRAUG canoe \\
\hline
\multirow{2}{*}{HI} & at lake LINDEMAN i had one canoe very good PETERBURGH canoe \\
 & at lake linderman i had one canoe A very good PEACHERBROROU canoe \\
\hline
\end{tabular}
\caption{\label{tab:transcripts} 
    Examples of transcription output of selected utterances from the Test set of Model-4 among all six L1s without a language model. Capitalized words indicate errors. We show samples from two speakers per accent.
}
\end{table*}


\subsection{Accent-Specific Models}
We evaluate the accent-dependent performance by fine-tuning our models on a single type of L1-specific accent at a time. 

    \textbf{Model-4 (speaker-dependent, accent-dependent):} The model is fine-tuned with Split-4 data to identify any accent-dependent training impact on downstream performance, as well as an upper bound on performance when the model is optimized for a single accent. In addition to evaluating on L2-ARCTIC Test, we test the model on L1-ARCTIC Test and LibriSpeech as a means to identify any degradation on L1 English data. 
    
    As Table~\ref{model-4} shows, while the multi-accent model (Model-1) outperforms Baseline-I for all six accents, all of the accent-specific models (Model-4 setting) outperform Model-1 on the Test\textsubscript{L2} setting despite the small amount of data (roughly five hours) used for fine-tuning each of the versions of Model-4. On average, Model-4 setting is two points WER better than Model-1. In addition, Model-4 type models (each of which is fine-tuned on one non-native accent) perform reasonably well on L1 data (Test\textsubscript{L1}, LS\textsubscript{Clean}, and LS\textsubscript{Other}). Further, large accent-specific variability is observed across different model types on Test\textsubscript{L2} ($SD$ = [$2.72-7.30$]), compared with native counterparts such as Test\textsubscript{L1} ($SD$ = [$0.20-0.43$]). An interesting result is the apparent difficulty difference between different accents ($HI$ and $KO$ easiest, $VI$ hardest), regardless of model types. We provide sample outputs from Model-4 in Table~\ref{tab:transcripts}. \par
    
    As shown in Table~\ref{model-4 zeroshot}, we also perform accent-wise zero-shot evaluation. Results of this set of experiments reveal an interesting pattern: while fine-tuning on a single accent generally benefits \textit{at least one other accent}, fine-tuning on the Hindi accent only benefits Hindi (the same accent) and hinders performance on \textit{all the other accents}. Figure~\ref{fig:model4-zeroshot1} and Figure~\ref{fig:model4-zeroshot2} illustrate this observation.
    
 \textbf{Model-5 (speaker-independent and accent-dependent):} This setup simulates a more realistic scenario where we target a single accent, without access to all speakers during development time. Thus, we use Split-5 data which mimics a speaker-independent setting. We cross-validate each L1 subset with one of the four speakers per fold. The hyper-parameters we use are those identified for Model-4. To evaluate the performance on each speaker, we conduct $24$ folds in total with one speaker removed at a time, and report the average and standard deviation of the four folds per each accent. 
 
 As Table~\ref{model-5} shows, speaker-dependent variability is small for Test\textsubscript{all} ($SD$ = [$0.11-0.38$]) but large for Test\textsubscript{zeroshot-speaker} ($SD$ = [$1.12-4.87$]). These results suggest that individual speaker's differences may play an important role in how much performance gain can be obtained by fine-tuning.\footnote{For those speakers whose TOEFL scores are known~\cite{zhao2018l2}, a strong negative correlation was observed between speaker-specific WERs of Baseline-I and speaker's TOEFL scores, $r$($8$) $ \approx -.77$, $p$ \textless $.01$.} 

\section{Conclusion}\label{sec:con}
We demonstrated potential of developing accent-independent and accent-dependent models that improve non-native speech recognition simply by fine-tuning the pre-trained wav2vec 2.0 model on a small amount of labeled data. Both the multi- and single-accent models improve performance on L2 English speakers. However, each accent benefits differently: results of the multi-accent, zero-shot experiments suggest that transfer learning on accent is possible and single-accent models improve the most for the target L2 accents.\par
As to future work, while we chose a language model-free setting to focus specifically on wav2vec 2.0's acoustic capacity, comparison with language model decoding would be a useful direction to explore as a way to gauge any further potential improvements a language model can bring. In addition, finding the optimal combination of accented speech datasets when there is no available dataset for a target accent (Model-3) may constitute another interesting direction. Finally, although we have offered a number of sample transcriptions from one of our models, a thorough error analysis on each experiment would help advance the research into improving ASR models for non-native English speakers. Since L2 English speakers have specific accent characteristics influenced by their native languages, an error analysis focused on each language as well as on groups or families of languages will likely aid effective model development. Future directions could also investigate different strategies for developing ASR systems for challenging languages such as Vietnamese. 

\bibliographystyle{acl_natbib}
\bibliography{anthology,acl2021}

\end{document}



\begin{table*}[htbp]
\centering
\begin{tabular}{lrrr}
\hline
\textbf{Models} & \textbf{Feature mask prob.} & \textbf{Feature mask len.} & \textbf{Timestep mask prob.} \\ \hline
Model-1 & 0.5 & 15 & 0.5 \\
Model-2 & 0.25 & 30 & 0.75 \\
Model-3 & 0.5 & 15 & 0.5 \\
Model-4 & 0.25 & 30 & 0.75 \\
Model-5 & 0.25, 0.5 & 15, 30 & 0.5, 0.75 \\ \hline
\end{tabular}
\caption{\label{tab:hyperparam masking} 
    Fine-tuning hyperparameters (masking). The parameter with multiple values means we set different value for different random seeds.}
\end{table*}

\begin{table*}[htbp]
\centering
\begin{tabular}{lrrrr}
\hline
 & \multicolumn{4}{c}{\textbf{Learning rate schedule}} \\ \cline{2-5} 
\textbf{Models} & \textbf{Warm-up} & \textbf{Constant} & \textbf{Decay} & \textbf{Constant} \\ \hline
Model-1 & 3e-5 & 3e-5 & 3e-5 & 1e-5 \\
Model-2 & 3e-5 & 3e-5 & 5e-6 & 2e-6 \\
Model-3 & 1e-5 & 1e-5 & 1e-5 & 5e-6 \\
Model-4 & 1e-5 & 1e-5 & 1e-5, 5e-6 & 2e-6, 5e-6 \\
Model-5 & 1e-5 & 1e-5 & 1e-5, 5e-6 & 2e-6, 5e-6 \\ \hline
\end{tabular}
\caption{\label{tab:hyperparam lr} 
    Fine-tuning hyperparameters (learning rates). To mimic the tri-state learning rate schedule, we control the learning rate over the warm-up, constant, and decay stages. The decay stage is followed by another constant stage that mimics \texttt{final\_lr\_scale} of the Fairseq's fine-tuning configuration. For \texttt{max\_step}, we pick from the set \textit{\{5000, 7000, 10000, 12500, 20000\}} for each stage. For \texttt{num\_warmup\_steps}, we pick from the set \textit{\{2000, 2500, 3000, 5000\}}. The \texttt{\{batch\_size\}} is set to $16$ for all models. The parameter with multiple values means we set different value for different random seeds.}
\end{table*}


\begin{table*}[]
\centering
\begin{tabular}{lrrr}
\hline
\textbf{Models} & \textbf{Feature mask prob.} & \textbf{Feature mask len.} & \textbf{Timestep mask prob.} \\ \hline
Model-1 & 0.5 & 15 & 0.5 \\
Model-2 & 0.25 & 30 & 0.75 \\
Model-3-VI & 0.5 & 15 & 0.5 \\
Model-3-ZH & 0.5 & 15 & 0.5 \\
Model-3-AR & 0.5 & 15 & 0.5 \\
Model-3-ES & 0.5 & 15 & 0.5 \\
Model-3-KO & 0.5 & 15 & 0.5 \\
Model-3-HI & 0.5 & 15 & 0.5 \\
Model-4-VI & 0.25 & 30 & 0.75 \\
Model-4-ZH & 0.25 & 30 & 0.75 \\
Model-4-AR & 0.25 & 30 & 0.75 \\
Model-4-ES & 0.25 & 30 & 0.75 \\
Model-4-KO & 0.25 & 30 & 0.75 \\
Model-4-HI & 0.25 & 30 & 0.75 \\
Model-5-VI & 0.25, 0.5 & 15, 30 & 0.5, 0.75 \\
Model-5-ZH & 0.25, 0.5 & 15, 30 & 0.5, 0.75 \\
Model-5-AR & 0.25, 0.5 & 15, 30 & 0.5, 0.75 \\
Model-5-ES & 0.25, 0.5 & 15, 30 & 0.5, 0.75 \\
Model-5-KO & 0.25, 0.5 & 15, 30 & 0.5, 0.75 \\ \hline
\end{tabular}
\caption{\label{tab:hyperparam masking - full)} 
    FULL VERSION. Fine-tuning hyperparameters (masking). The parameter with multiple values means we set different value for different random seeds.}
\end{table*}

\begin{table*}[]
\centering
\begin{tabular}{lrrrrr}
\hline
 & \multicolumn{5}{c}{\textbf{Learning rate schedule}} \\ \cline{2-6} 
\textbf{Models} & \textbf{Warm-up} & \textbf{Warmup steps} & \textbf{Constant} & \textbf{Decay} & \textbf{Constant} \\ \hline
Model-1 & 3e-5 & 2500 & 3e-5 & 3e-5 & 1e-5 \\
Model-2 & 3e-5 & 5000 & 3e-5 & 5e-6 & 2e-6 \\
Model-3-VI & 1e-5 & 2000 & 1e-5 & 1e-5 & 5e-6 \\
Model-3-ZH & 1e-5 & 2000 & 1e-5 & 1e-5 & 5e-6 \\
Model-3-AR & 1e-5 & 2000 & 1e-5 & 1e-5 & 5e-6 \\
Model-3-ES & 1e-5 & 2000 & 1e-5 & 1e-5 & 5e-6 \\
Model-3-KO & 1e-5 & 2000 & 1e-5 & 1e-5 & 5e-6 \\
Model-3-HI & 1e-5 & 2000 & 1e-5 & 1e-5 & 5e-6 \\
Model-4-VI & 1e-5 & 3000 & 1e-5 & 5e-6 & 2e-6 \\
Model-4-ZH & 1e-5 & 2000 & 1e-5 & 1e-5 & 5e-6 \\
Model-4-AR & 1e-5 & 2000 & 1e-5 & 1e-5 & 5e-6 \\
Model-4-ES & 1e-5 & 2000 & 1e-5 & 5e-6 & 2e-6 \\
Model-4-KO & 1e-5 & 2000 & 1e-5 & 1e-5 & 5e-6 \\
Model-4-HI & 1e-5 & 2000 & 1e-5 & 5e-6 & 2e-6 \\
Model-5-VI & 1e-5 & 3000 & 1e-5 & 5e-6 & 2e-6 \\
Model-5-ZH & 1e-5 & 2000 & 1e-5 & 1e-5 & 5e-6 \\
Model-5-AR & 1e-5 & 2000 & 1e-5 & 1e-5 & 5e-6 \\
Model-5-ES & 1e-5 & 2000 & 1e-5 & 5e-6 & 2e-6 \\
Model-5-KO & 1e-5 & 2000 & 1e-5 & 1e-5 & 5e-6 \\ \hline
\end{tabular}
\caption{\label{tab:hyperparam lr - full} 
    FULL VERSION. Fine-tuning hyperparameters (learning rates). To mimic the tri-state learning rate schedule, we control the learning rate over the warm-up, constant, and decay stages. The decay stage is followed by another constant stage that mimics \texttt{final\_lr\_scale} of the Fairseq's fine-tuning configuration. For \texttt{max\_step}, we pick from the set \textit{\{5000, 7000, 10000, 12500, 20000\}} for each stage. For \texttt{num\_warmup\_steps}, we pick from the set \textit{\{2000, 2500, 3000, 5000\}}. The \texttt{\{batch\_size\}} is set to $16$ for all models. The parameter with multiple values means we set different value for different random seeds.}
\end{table*}